# Programmable photonic time circuits for highly scalable universal unitaries

Xianji Piao[1], Sunkyu Yu[2†] and Namkyoo Park[1*]

[1]Photonic Systems Laboratory, Department of Electrical and Computer Engineering, Seoul National University, Seoul 08826, Korea

[2]Intelligent Wave Systems Laboratory, Department of Electrical and Computer Engineering, Seoul National University, Seoul 08826, Korea

E-mail address for correspondence: [†]sunkyu.yu@snu.ac.kr, [*]nkpark@snu.ac.kr

**Abstract**

Programmable photonic circuits (PPCs) have garnered substantial interest in achieving deep learning accelerations and universal quantum computations. Although photonic computation using PPCs offers critical advantages, including ultrafast operation, energy-efficient matrix calculation and room-temperature quantum states, its poor scalability impedes the integration required for industrial applications. This challenge arises from the temporally one-shot operation using propagating light in conventional PPCs, which leads to the light-speed increase of device footprints. Here we propose a concept of programmable photonic time circuits, which employ time-cycle-based computations analogous to the gate cycling in the von Neumann architecture and quantum computation. As a building block, we develop a reconfigurable SU(2) time gate composed of two resonators, which have tunable resonances and are coupled through time-coded dual-channel gauge fields. We demonstrate universal U($N$) operations with high fidelity using the systematic assembly of the SU(2) time gates, achieving improved scalability from $O(N^2)$ to $O(N)$ in both the footprint and gate number. This result opens a pathway to industrial-level PPC implementation in very large-scale integration.

# Introduction

A programmable photonic circuit (PPC) is a versatile platform for neuromorphic and quantum computation[1,2] by supplying run-time tunability in addition to the inherent advantages of photons as signal carriers—ultrafast propagation, broad bandwidth, robust quantum states[3] and energy-efficient matrix calculation[4,5]. One of the critical goals of PPCs is to achieve a universal unitary operation U($N$) with reconfigurability, which has been employed to realize trainable weight matrices in wave neural networks[4] and programmable quantum gates for general-purpose linear optical quantum computation (LOQC)[6,7]. The conventional approach to realize U($N$) PPCs is to utilize a set of universal SU(2) optical gates, which can be implemented with two beam splitters or interferometers, and two phase shifters[2]. The arrangements and device parameters of the gates are determined systematically for a given unitary matrix[8,9].

In realizing high-$N$ unitary PPCs for photonic deep learning with a large number of neurons and for multi-qubit LOQC with high-dimensional quantum gates, two major hurdles remain: fidelity and footprint. First, as the size of PPCs increases with higher $N$, the need for high-fidelity platforms becomes more crucial due to increasing manufacturing errors and thermal noises. Therefore, consistent efforts have been made to improve the fidelity of high-$N$ PPCs using self-calibration[10], optomechanics[11], or circuit pruning[12]. Another critical challenge for high-$N$ unitary PPCs, which can exacerbate the first hurdle, is the poor scalability of circuit footprints. This difficulty stems from the underlying design philosophy of traditional PPCs: U($N$) operations of propagating light, which results in an increasing optical path length. Consequently, the two-dimensional (2D) footprint of the PPC is proportional to the product of the optical path length and the number of optical channels, and is also directly related to the number of employed optical gates.

In traditional algorithms[8,9], U(N) circuits exhibit $O(N^2)$ scaling for both the 2D footprint and the number of optical elements.

Significant efforts have been conducted to resolve this scaling issue by exploiting advanced platforms or further physical axes. First, recent efforts have tried to realize integrated PPCs using subwavelength optics[13] or diffractive elements[14,15]. However, the broadening in the momentum space of photons, such as the emergence of backscattering, hinders unitary operations at the bounded direction, which degrades the fidelity or scalability of PPCs. Second, an elegant approach to utilize the frequency synthetic dimension[16] has enabled $O(N)$ scaling in the PPC device size. Yet, this approach requires an N number of multiwavelength-coded light sources and detectors, restricting the allowed values of N with the free spectral ranges of the sources and device bandwidths. In this context, it still remains challenging to realize high-N unitary PPCs with scalability, for example, by exploiting single wavelength operations. Notably, when we revisit various forms of computations along the temporal axis—the fetch-execute cycles in the von Neumann architecture[17], the qubit-gate cycles in quantum computing[18] and the synaptic plasticity of the brain[19]—we can envisage the utilization of a temporal degree of freedom for PPCs.

Here we develop the PPC platform achieving $O(N)$ scaling for universal unitary operations with the single wavelength light sources and detectors. Inspired by the space-time duality, which has raised the concepts of photonic time crystals[20,21], time disorder[22,23] and time diffraction[24], we propose programmable photonic "time" circuits (PPTCs) composed of coupled resonators to replace the optical path length with the temporal field evolution. As a unit element for composing U(N) operations, we devise an SU(2) time gate, which can apply reconfigurable SU(2) operations to stored light. By employing the SU(2) time gates to the Clements design, we demonstrate the realization of random Haar matrices and quantum Fourier transforms with $O(N)$ scaling for both the footprint and the number of optical gates. This result provides an improved platform for the integration toward large-scale photonic deep learning accelerators and quantum computations.

## Results

**Concept of programmable photonic time circuits**

To introduce the concept of PPTCs, we compare the realizations of the unitary matrix $U_N \in U(N)$ using conventional space-domain ($x$) PPCs (Fig. 1a) and our time-domain ($t$) PPTCs (Fig. 1b). Each circuit is composed of its corresponding SU(2) unit operations using a propagating (Fig. 1c) or resonance (Fig. 1d) mode of light. $U_N$ is implemented through diagonalization using the nulling process[9] (Fig. 1e): the cascaded multiplication of the inversely designed U($N$) unit matrices that apply SU(2) operations to two specific channels in order to set the off-diagonal elements of $U_N$ to be zero. Among various types of nulling processes, we employ the Clements design[9], which provides more integrated and loss-tolerant unitary circuits than those of the original Reck design[8] in space-domain PPCs.

In realizing $U_N$ with PPCs or PPTCs, $N(N-1)/2$ SU(2) operations are required to handle all the off-diagonal elements of $U_N$. The Clements design facilitates a symmetric arrangement of the SU(2) gates[9], which in turn enables the execution of an average of $(N-1)/2$ SU(2) operations simultaneously at each stage along the $x$-axis (Fig. 1a). Therefore, the conventional PPCs lead to $O(N^2)$ scaling on their 2D footprints, requiring $N(N-1)$ passive Mach-Zehnder interferometers (MZIs) and $N^2$ active phase shifters when using the SU(2) gates in Fig. 1c. These numbers are evidently incompatible with large-scale deep neural networks or noisy intermediate-scale quantum computing (NISQ). For example, when considering an example of the spatial SU(2) gates[11] having the 2D footprint of about 0.5 mm$^2$, a unitary matrix for $10^3$ photonic neurons or 10 qubits requires the circuit of ~0.5 m$^2$ size with about 10$^6$ pairs of passive and active photonic elements. This poor scalability prohibits the use of PPCs for large-scale problems in deep learning and quantum computation in the near future.

To tackle this hurdle, we propose the PPTC by exploiting space-time duality[20-24] in photonics, targeting the integrated computation in the time domain. In this proposal, each photonic channel is defined with light stored inside a resonator. The use of resonances replaces the *x*-axis optical path length (Fig. 1a) with the *t*-axis evolution time (Fig. 1b), leading to $O(N)$ scaling in the spatial circuit footprint. As a unit element, we devise an SU(2) time gate (Fig. 1d) of which the operation principle will be discussed in the next section. Because the gate can apply reconfigurable SU(2) operations to the standing waves stored inside resonators, the PPTC allows for $O(N)$ scaling also for the number of optical gates. Table 1 shows the scalability of the spatial PPCs using Reck and Clements designs and our PPTC.

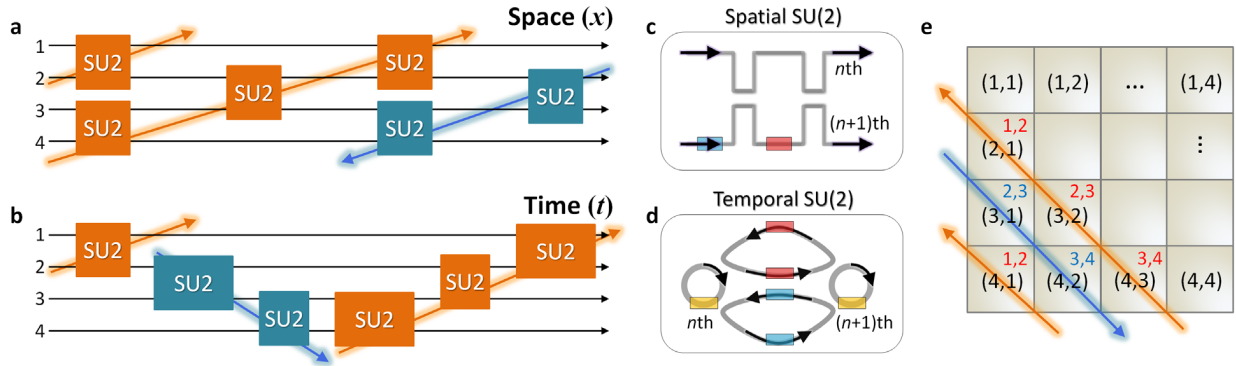

**Fig. 1. PPTCs for universal unitaries with $O(N)$ scalability. a,b,** U(4) implementations using the conventional PPC (**a**) and the PPTC (**b**). Black arrows denote the evolution of optical modes along either the spatial (*x*-) or temporal (*t*-) axis. Each coloured box labelled 'SU2' indicates the SU(2) optical gate for two adjacent channels, while the horizontal length of the box represents the spatial or temporal footprint of the gate. **c,d,** The building blocks for the SU(2) operations between the *n*th and (*n*+1)th channels in the PPC (**c**) and the PPTC (**d**). The PPC building block—the SU(2) space gate—consists of two MZIs (grey lines) and two tunable phase shifters (coloured boxes) that employ propagating modes (**c**). The proposed PPTC building block—the SU(2) time gate—is composed of two resonators (grey circles) coupled via two zero-field waveguide loops (grey curved triangles), which support travelling-wave resonance modes (**d**). Tunable phase shifters are applied to the resonators (yellow boxes; **d**) and waveguides or waveguide loops (red and blue boxes; **c** and **d**). Black arrows in **c** and **d** denote the direction of wave propagations. **e,** An example of the nulling process using the Clements design[9] for the diagonalization of $U_4$. The pairs $(p,q)$ (black) and $r,s$ (red and blue) denote the matrix element index and the channels connected through the SU(2) gate for nulling $(p,q)$, respectively ($p$, $q$, $r$ and $s$ are the integers among 1, 2, 3 and 4). Coloured arrows and boxes in **a**, **b**, and **e** represent two sub-processes of diagonalization, which will be discussed later.

**Table 1. Scalabilities of PPCs and PPTCs.** The footprint of the phase shifter for a diagonal matrix is assumed to be half of each SU(2) gate in the Reck or Clements PPCs.

|  | **Reck PPC[8]** | **Clements PPC[9]** | **PPTC** |
|---|---|---|---|
| **Passive elements** | $N(N-1)$ MZIs | $N(N-1)$ MZIs | $N$ resonators and $2(N-1)$ waveguide loops |
| **Active elements** | $N^2$ phase shifters | $N^2$ phase shifters | $5N-4$ phase shifters |
| **2D footprint** | $N \times (2N - 3 + 1/2)$ | $N \times (N + 1/2)$ | $N$ |
| **SU(2) per gate** | 1 | 1 | $N/2$ |

**SU(2) time gates**

In replacing the *x*-axis propagation with the *t*-axis evolution, one of the critical issues is achieving a temporal reproduction of wave behaviours in spatially varying structures. For example, the temporal analogy of the SU(2) space gate is nontrivial. Upon initial observation, one might presume the use of the diatomic resonators that possess dynamically controlled resonant frequencies and coupling coefficients as the analogies of the phase shifts and MZI couplings, respectively. However, the dynamical engineering of coupling is technically challenging because the coupling coefficient is primarily determined by the rigid physical distance between resonators.

Such an obstacle exacerbates the challenge of realizing reconfigurable time-domain SU(2) operations seamlessly covering the entire Bloch sphere.

To address this challenge, we propose the PPTC composed of reconfigurable and universal SU(2) time gates. The PPTC is the lattice of coupled resonators, which support the pseudo-spin modes of clockwise ($\sigma = +1$) and counter-clockwise ($\sigma = -1$) wave circulations[25] at a given frequency. The neighbouring resonators are coupled through two zero-field waveguide loops[25,26] to achieve the evanescent coupling with dual-channel gauge fields (Fig. 2a; Supplementary Note S1). First, the effective coupling strength is determined by the decay rate $1/\tau$ of both pseudo-spin modes to a waveguide loop. Second, for the upper (U) and lower (L) waveguide loops between the $m$th and $n$th resonators, we apply the time-varying phase differences between two waveguide arms, as $\pm\xi_{mn}^{U}(t)$ and $\pm\xi_{mn}^{L}(t)$ (Fig. 2a). These phase differences drive dynamical dual-channel gauge fields $\xi_{mn}^{U}(t)$ and $\xi_{mn}^{L}(t)$ acquired along the paths from the $n$th to $m$th sites, which have opposite signs for pseudo-spin modes. The tight-binding Hamiltonian of the PPTC lattice is then (Supplementary Note S1):

$$H = -\sum_{m,\sigma}\left[\omega_0 + \Delta\omega_m(t)\right]a_{m\sigma}^{\dagger}a_{m\sigma} - \frac{1}{2\tau}\sum_{\langle m,n\rangle,\sigma}\left[\left(e^{-i\sigma\xi_{mn}^{U}(t)} + e^{-i\sigma\xi_{mn}^{L}(t)}\right)a_{m\sigma}^{\dagger}a_{n\sigma} + \text{H.c.}\right], \quad (1)$$

where $\omega_0$ is the reference resonant frequency, $\Delta\omega_m(t)$ is the time-varying resonance perturbation of the $m$th resonator, $a_{m\sigma}^{\dagger}$ and $a_{m\sigma}$ are the creation and annihilation operators for the $\sigma$ pseudo-spin mode at the $m$th site, respectively, the pair $\langle m,n\rangle$ is the neighbouring indices for the coupled sites, and H.c. denotes the Hermitian conjugate. In the lattice described by Eq. (1), the unit cell composed of diatomic resonators (Fig. 2a) operates as the SU(2) time gate.

Because the fundamental strategy of achieving a high-degree $U_N$ is to decompose $U_N$ into a set of SU(2) operations[8,9], we examine a range of SU(2) operations accessible with a SU(2) time gate, by setting $\xi_{mn}^{U,L}(t) = \xi^{U,L}(t)$, $\Delta\omega_m(t) = +\Delta\omega(t)$, and $\Delta\omega_n(t) = -\Delta\omega(t)$ in Eq. (1). Focusing on the $\sigma = +1$ pseudo-spin mode of which the field amplitude in the $p$th resonator is $\psi_p$, we introduce the spinor state $\Psi = [\psi_m, \psi_n]^T$. The spinor satisfies the governing equation $id\Psi/dt = H_S\Psi$, where the dynamical Hamiltonian $H_S$ is (Supplementary Note S2)

$$H_S = -\omega_0\sigma_0 - \frac{1}{2\tau}\left[\cos\xi^U(t) + \cos\xi^L(t)\right]\sigma_x - \frac{1}{2\tau}\left[\sin\xi^U(t) + \sin\xi^L(t)\right]\sigma_y - \Delta\omega(t)\sigma_z, \quad (2)$$

where $\sigma_0$ and $\sigma_{x,y,z}$ are the identity matrix and Pauli matrices, respectively.

Although Eqs. (1) and (2) generally results in nonlinear dynamics due to time-varying system parameters, we impose the linearized picture on $id\Psi/dt = H_S\Psi$ to gain better insight into the SU(2) operations and the following unitary approximation. The linearization is achieved by assuming the digital modulation of the system parameters, which leads to the constant values of the gauge fields and resonance perturbation, as $\xi^{U,L}(t) = \xi^{U,L}$ and $\Delta\omega(t) = \Delta\omega$, at a given temporal range. When we represent the spinor state with the Stokes vector[27] $\mathbf{S} = [S_x, S_y, S_z]^T$ where $S_j = \Psi^\dagger\sigma_j\Psi$ ($j = x, y,$ and $z$), the geometrical evolution of $\mathbf{S}$ on the Bloch sphere is governed by $d\mathbf{S}/dt = \mathbf{S} \times \mathbf{B}$, where the pseudo-magnetic field $\mathbf{B}$ for the spinor is (Supplementary Note S3)

$$\mathbf{B} = \frac{2}{\tau}\begin{bmatrix}(\cos\xi^U + \cos\xi^L)/2 \\ (\sin\xi^U + \sin\xi^L)/2 \\ \Delta\omega\tau\end{bmatrix}. \quad (3)$$

Equation (3) demonstrates that the evolution of the spinor state in the SU(2) time gate is analogous to the Larmor precession of a magnetic moment[28] under the pseudo-magnetic field $\mathbf{B}$, which can be freely manipulated with the system parameters of the gate. According to this analogy, Eq. (2) can be rewritten as $H_S = -\omega_0\sigma_0 - \mathbf{B}\cdot\boldsymbol{\sigma}/2$ with the linearized picture, where $\boldsymbol{\sigma} = \mathbf{e}_x\sigma_x + \mathbf{e}_y\sigma_y + \mathbf{e}_z\sigma_z$ is the Pauli vector. During the linearized temporal range of the length $t$, we achieve the spinor precession on the Bloch sphere about $\mathbf{B} = \mathbf{e}_B B_0$, which corresponds to the SU(2) rotation operation $R_\mathbf{B}(t)$ of $\Psi$ (Supplementary Note S4), as[29]

$$R_{\mathbf{B}}(t) = e^{i\omega_0 t}\left[\left(\cos\frac{B_0 t}{2}\right)\sigma_0 + i\left(\sin\frac{B_0 t}{2}\right)\mathbf{e_B}\cdot\boldsymbol{\sigma}\right]. \quad (4)$$

According to Eq. (4), the universal SU(2) rotation can be obtained by altering the dual-channel gauge fields $\xi^{U,L}$, the resonance perturbation $\Delta\omega$, and the duration of the temporal range $t$.

For the systematic realization of a high-degree $U_N$, we define two orthogonal system states: the state of the even-parity gauge fields without resonance perturbations (Fig. 2b with $\xi^U = \xi^L = \xi$ and $\Delta\omega = 0$) and the state of the odd-parity and quarter-wave gauge fields with resonance perturbations (Fig. 2c with $\xi^U = -\xi^L = \pi/2$, while allowing for $\Delta\omega \neq 0$). The even and odd states correspond to the rotations of a spinor state about $\mathbf{B}$ on the $xy$ plane (Fig. 2d) and along the $z$-axis (Fig. 2e), respectively, which enable the complete coverage of the Bloch-sphere surface. The magnitude of each rotation can be controlled by changing the temporal duration $t$ (Supplementary Note S5).

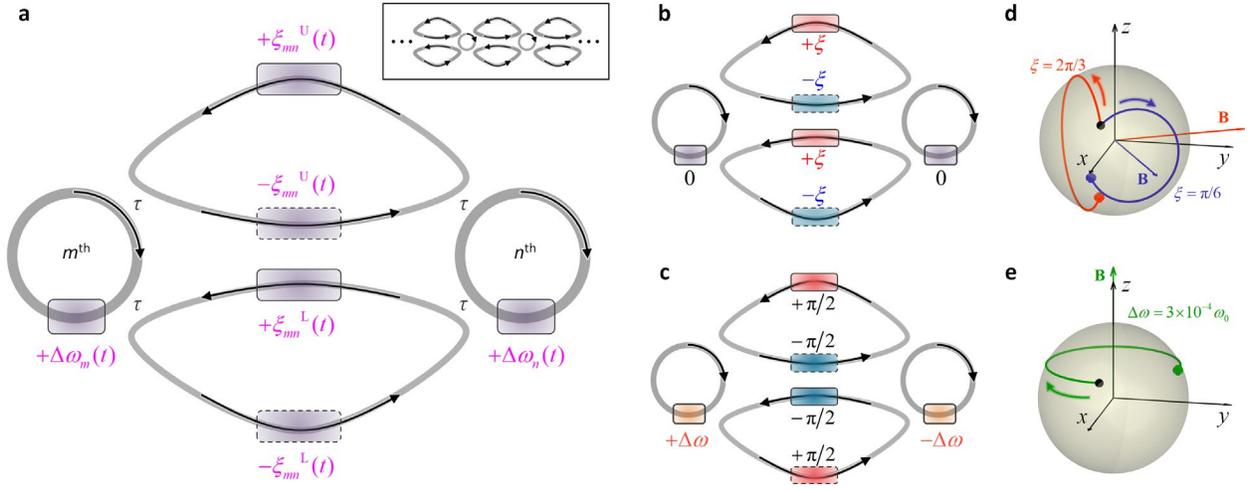

**Fig. 2. SU(2) time gate for PPTCs. a,** Schematic diagram of the SU(2) time gate composed of two resonators (circles) and two waveguide loops (curved triangles). The shaded boxes represent the tunable phase shifters for the resonance perturbations $\Delta\omega_{m,n}(t)$ and the dual-channel gauge fields $\xi_{mn}^{U,L}(t)$. The inset in **a** represents the PPTC composed of the SU(2) time gates. **b,c,** Two orthogonal system states with the even-parity (**b**) and odd-parity (**c**) gauge fields. Black arrows in **a-c** denote the direction of wave circulations. **d,e** The spinor evolutions on the Bloch sphere for the even-parity states (**d**) with $\xi = 2\pi/3$ (red) and $\xi = \pi/6$ (blue), and the odd-parity state (**e**) with $\Delta\omega = 3\times10^{-4}\omega_0$. In **d** and **e**, the plotted temporal range is from 0 to $0.8\pi\tau$ where the black dots denote the initial states.

**U(N) operations**

To decompose a high-degree $U_N$ matrix with the proposed SU(2) time gates in Figs. 2b and 2c, we extend the Clements design[9] for conventional PPCs to our PPTCs (Supplementary Note S6). Same as the Clements design, the proposed method for PPTCs is based on the nulling of the super- and sub-diagonals of $U_N$ by sequentially applying $U_N(T_m^l)^\dagger$ for the $(l, m)$ element nulling and $(T_m^l)^\dagger U_N$ for the $(m+1, l)$ element nulling, where $T_m^l \in U(N)$ leads to the SU(2) operation between the $m$th and $(m+1)$th channels and preserves the remaining channels. The entire nulling process derives the decomposed realization of $U_N$, as follows (Supplementary Note S6):

$$U_N = \left[\prod_{(l_F, m_F)'} T_{m_F}^{l_F}\right] D \left[\prod_{(l_B, m_B)'} T_{m_B}^{l_B}\right], \qquad (5)$$

where the sequences of the index pairs $(l_F, m_F)'$ and $(l_B, m_B)'$ are determined by an order of the nulling processes, and $D$ is the resulting diagonal matrix after the entire nulling.

The $U_N$ decomposition in Eq. (5) with the linearized picture for Eq. (2) leads to the digitized temporal evolutions of the system parameters: the dual-channel gauge fields between the $p$th and $(p+1)$th resonators $\xi^{U,L}_{(p,p+1)}(t)$, and the resonance perturbation of the $q$th resonator $\Delta\omega_q(t)$ ($1 \leq p \leq (N-1)$, $1 \leq q \leq N$). The constant values of each system parameters with the duration time $t$ determine a specific sub-matrix $T_m^l$ or $D$. Therefore, PPTCs allow for achieving universal unitaries through the time-coded digital modulation of a series of SU(2) time gates, performing the cycle-based computations along the temporal axis instead of the temporally one-shot, spatial computation in conventional PPCs. We also note that the duration time $t$ is proportional to the lifetime $\tau$ of each resonator (Supplementary Note S6). In consequence, stronger coupling between resonators with smaller $\tau$ enables more rapid calculation of a unitary matrix.

As the first example of the PPTC operation, we examine the realization of the unitary matrix $U^{QFT} \in U(N)$ for the quantum Fourier transform (QFT)[29]. In applying the time-coded modulations to each resonator and waveguide loop for achieving $U^{QFT}$, we assume the first-order low-pass filtering to the modulation signals with the cutoff frequency $\omega_c$ to reflect the response time of refractive index changes (see Methods). For the low-pass-filtered modulations, we conduct the time-domain analysis of the system described by Eq. (1) using the sixth-order Runge-Kutta method[30] (see Methods for time-domain analysis).

Figure 3 shows the two-qubit QFT ($N = 2^2$) achieved with the PPTC. Figure 3a presents the low-pass-filtered modulation of the resonators and waveguide loops with the cutoff frequency $\omega_c = \omega_0 / 100$, where $\omega_0$ is the operation frequency of the PPTC. Through the time-coded modulations, the amplitude and phase of the field inside each resonator are tailored as the nulling unitaries $T_m^l$ and the diagonal $D$ are multiplied according to Eq. (5) (Fig. 3b). Figures 3c and 3d show an example of the input and output of the two-qubit QFT. As shown in Fig. 3d, the obtained output (solid lines) closely resembles the solution (dashed lines), although the small deviation, which likely originates from the low-pass filtering requires further quantitative analysis, as described in the next section.

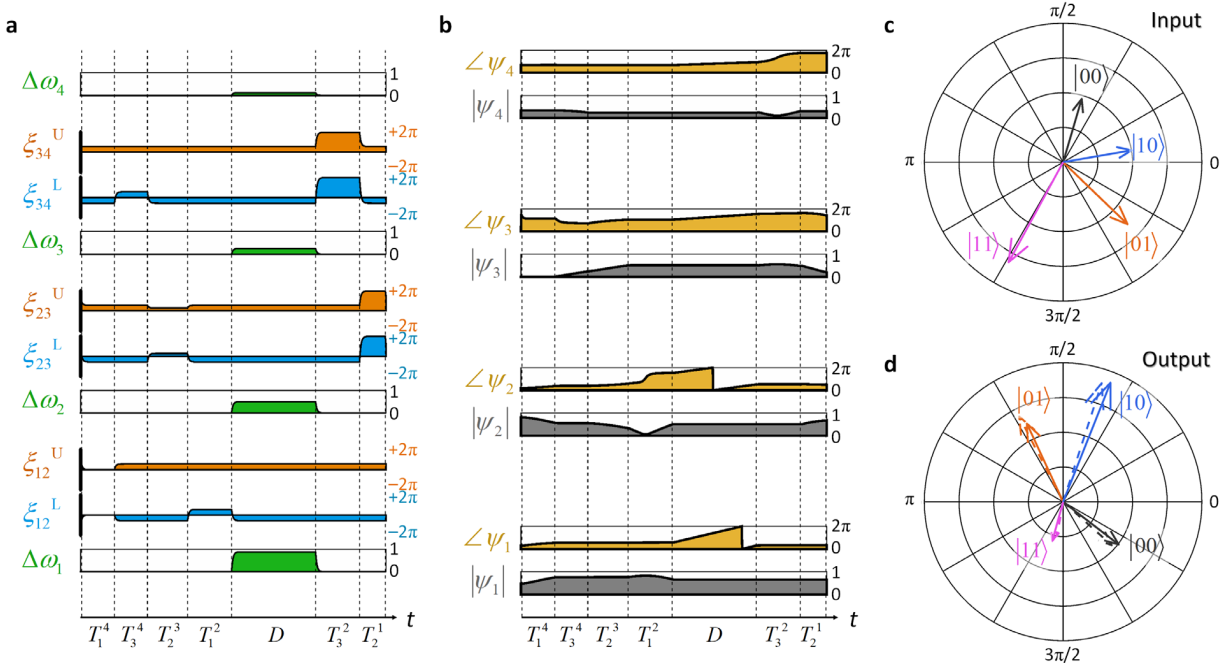

**Fig. 3. Two-qubit QFT PPTC. a,** The time-coded modulations of the resonance perturbations $\Delta\omega_q$ and the gauge fields in the waveguide loops $\xi^{U,L}_{(p,p+1)}$ for the two-qubit QFT $U^{QFT}$. The signals undergo low-pass-filtering with the cut-off frequency $\omega_c = \omega_0 / 100$. The duration time of each temporal cell is determined by the elements of $U^{QFT}$ according to Supplementary Note S6. **b,** The corresponding temporal evolutions of the state vectors (modulus and angle of $\psi_q$) at each step of nulling processes. **c,d,** An example of the input (**c**) and output (**d**) of the two-qubit QFT operations. The resonance fields that represent four qubit states $|00\rangle$, $|01\rangle$, $|10\rangle$ and $|11\rangle$ are expressed in the complex plane. In **d**, the solid and dashed lines describe the PPTC output and the solution, respectively. The lifetime of the resonators is set to be $\tau = 500 \times (2\pi / \omega_0)$.

**Fidelity and measurement**

Extending an operation of the QFT example in Fig. 3, we investigate the fidelity of the PPTC for the QFT and universal unitaries. In examining universal unitaries, we apply uniform sampling of the U(N) group with the Haar measure[31], generating K realizations of random Haar matrices $U_k^{Haar}$ ∈ U(N) (k = 1, 2, …, K). Notably, when we attempt to realize a $U^{QFT}$ or $U_k^{Haar}$ using the PPTC, the time-coded digital modulations according to Eq. (5) generally result in a nonunitary operation because the governing equation of the PPTC in Eq. (1) is both nonlinear and non-Hermitian. Therefore, we develop the stochastic model for evaluating the PPTC fidelity (see Methods for details). The model utilizes $L$ number of the effective matrix operations $\{V^{QFT,l}\}$ for $U^{QFT}$ or $\{V_k^{Haar,l}\}$ for $U_k^{Haar}$ ($l = 1, 2, …, L$) obtained from the relationship between $M$ random inputs and their corresponding outputs calculated by the Runge-Kutta method, where $L, M \gg N$. By employing the fidelity[12] comparing the effective $N \times N$ matrix $V$ to $U$:

$$F(U,V) = \frac{2\text{Re}\left[\text{Tr}(V^\dagger U)\right]}{N + \text{Tr}(V^\dagger V)}, \qquad (6)$$

the PPTC fidelities for the QFT and Haar matrices are obtained as $F^{QFT} \equiv \langle F(U^{QFT}, V^{QFT,l})\rangle_l$ and $F_k^{Haar} \equiv \langle F(U_k^{Haar}, V_k^{Haar,l})\rangle_l$, respectively, where $\langle…\rangle_l$ denotes the ensemble average for the realizations with the indices $l$.

For a thorough investigation into the effect of the response time of refractive index changes, which results in defective modulations, we examine the fidelities $F^{QFT}$ and $F_k^{Haar}$ for different low-pass filtering bandwidths in Fig. 4a. We compare unitary operations for $N = 4$ (orange, 2 qubits) and $N = 8$ (blue, 3 qubits) in realizing random Haar matrices (circles and error bars for $K = 20$ realizations) and QFT unitary matrices (triangles). The result shows that higher cutoff frequencies $\omega_c$, which provide the modulations closer to the ideal design, lead to better fidelities. Although the importance of the modulation bandwidth $\omega_c$ becomes more significant as $N$ increases, three-qubit ($N = 8$) PPTCs achieve $F^{Haar} \geq 0.95$ for $\omega_c \geq 0.006\omega_0$, where $F^{Haar} \equiv \langle F_k^{Haar}\rangle_k$ for the averaged fidelity with the realizations of different $U_k^{Haar}$. It is important to note that the decrease in fidelity originates from our digital modulation scheme, specifically the deviation from the designed square-wave modulations due to low-pass filtering (Fig. 3a). Various techniques in digital signal processing[32], such as predistortion, can be directly applied to compensate for this fidelity degradation.

For practical implementation, it is also necessary to devise a measurement setup for the standing waves stored inside the coupled resonators of the PPTC. Similar to conventional approaches applied to coupled resonator lattices[25,26], we utilize a probe waveguide coupled to each resonator with an identical lifetime $\tau_e$ (Fig. 4b), which provides separate optical paths for the incident ($\varphi_n^+$) and scattering waves ($\varphi_n^-$) (see Methods for time-domain analysis). When we assume the excitation of temporally bound incident pulses to the coupled resonators and the normalization of the scattering power $\Sigma_n |\varphi_n^-|^2$, the probe waveguides have little to no impact on the fidelity of the PPTC (Supplementary Note S7). Instead, the relative magnitude between the internal and probe couplings characterized by $\tau_e/\tau$ determines the measured scattering power. Figure 4c shows the normalized scattering powers in the QFT PPTCs, as functions of $\tau_e/\tau$. We apply the Dirac-delta-function-like excitations of 100 normalized random inputs through the arrays of $\varphi_n^+$. When $\tau_e \gg \tau$, the probe waveguide is too weakly coupled to the PPTC to excite sufficient input to coupled resonators. In contrast, when $\tau_e \ll \tau$, the excited fields decay too rapidly during the application of the target unitary operations. Because the relationship between the input and output is a function of $N$, the competition between the wave excitation and decay results in an $N$-dependent optimum point in $\tau_e/\tau$ (Fig. 4c: orange and blue). We also note that because the scattering power is determined solely by $\tau_e$ regardless of $U_N$, we can achieve unitary operations deterministically by applying the normalization to the scattering power.

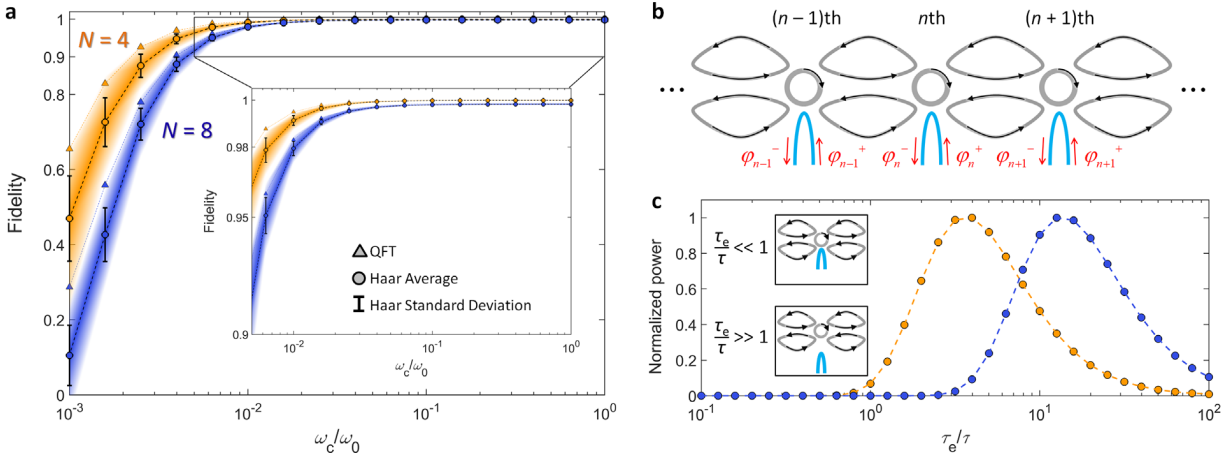

**Fig. 4. Fidelity and measurement of PPTCs. a,** The fidelities of the PPTCs as functions of the cutoff frequency $\omega_c$ in the low-pass-filtering: $F^{QFT}$ (triangles) and $F^{Harr}$ (circles) with $N = 4$ (orange, 2 qubits) and $N = 8$ (blue, 3 qubits). The inset in **a** presents the range of the fidelity larger than 0.9. Circles and error bars show the average and standard deviation of 20 random realizations at each $\omega_c$, respectively. The numbers of the random inputs for estimating $V^{QFT,l}$ or $V_k^{Haar,l}$ and the number of the random samplings for estimating effective matrix operations are set to be $M = 100$ and $L = 100$, respectively. **b,** Schematic diagram of the measurement setup for the PPTCs, employing probe waveguides (blue) to resonators. Red arrows denote the incident ($\varphi_n^+$) and scattering ($\varphi_n^-$) waves. **c,** Scattering power as a function of the lifetime $\tau_e$ to each probe waveguide. All the other parameters are the same as those in Fig. 3 and Methods.

## Discussion

Beyond the scalability comparison of a conventional PPC and our PPTC (Table 1), an approximate estimation of the SU(2) gate footprint in each circuit allows for addressing the integration issue more concretely. Consider the ~200 × 100 μm² footprint of the PPC SU(2) gate[4] and the ~50 × 50 μm² unit cell composed of coupled resonators linked by a waveguide loop without phase shifters[33]. If we assume the SU(2) time gate having the similar phase shifters with that of the PPC (approximately 35 μm length), the SU(2) time gate can be implemented within the footprint of the PPC SU(2) gate. Therefore, the PPC and PPTC fall within the large-scale integration (LSI) to very-large-scale integration (VLSI) regime with $10^3 \sim 10^4$ gates/mm$^2$ when focusing solely on the gate number. However, as demonstrated by its $O(N^2)$ scalability, the SU(2) gates in the PPC must be intertwined with each other to express universal unitaries, which limits the possible integration of the PPC to only medium-scale integration (MSI). Therefore, utilizing the temporal axis in the PPTC paves the way for a photonic VLSI matrix calculation accelerator.

Another intriguing opportunity offered by the PPTC is the potential to decompose U($N$) using higher-degree SU($M$) gates, where $2 < M < N$. When developing 2D planar integrated circuits, one axis of the spatial PPC corresponds to the stacking direction of the SU($M$) gates for achieving U($N$). Therefore, the coupling at a specific position is limited to occurring between two waveguides ($M = 2$) due to negligible long-range coupling with the remaining axis. This limitation underlies the operation principle of most existing U($N$) decomposition processes using SU(2) gates[8,9]: nulling a matrix element per step. In contrast, the PPTC can be implemented with 2D coupled-resonator lattices having different connectivity from Euclidean[34] to non-Euclidean[26,35] geometry. Such higher-dimensional configurations allow for higher-degree SU($M$) gates, which can reduce the number of nulling processes analogous to using qudits in quantum computation[36].

Despite the listed potentials, substantial hurdles must be overcome to realize high-$N$ PPTCs. Although the fidelity issue concerning the modulation speed can be resolved with the compensation techniques in digital signal processing, there still remains a trade-off relation between the intrinsic quality (Q-) factor of resonators and the necessary modulation speed. Because the sequence of the SU(2) time gates should be achieved during the flight time of light inside resonators, the Q-factor of the resonators determines the lower bound of the modulation speed. Considering the standard Q-factor (Q > $10^8$) in all-waveguide optical resonators, unitary operations need to be implemented during about 10 ns for the coupling Q that is 1/10 of the intrinsic

Q. Because U(*N*) operations require $N(N-1)/2 + 1$ steps of modulations, the necessary modulation speed for U(*N*) becomes about ($N^2$ / 10) GHz in the telecom wavelength. Although the electro-optical modulation[37] near 10 GHz will enable unitary operations for $N \sim 10$, applying SU($M > 2$) decomposition scheme, ultrahigh-Q resonators[38], the on-chip integration of gain media[39] and highly-sensitive photodetection[40] will dramatically alleviate this restriction. The use of on-chip coherent detection[41] will also allow for performing the sub-operations $T_m^l$ and $D$ of Eq. (5) in stages with much more alleviated modulation speed.

In conclusion, we proposed a new platform for achieving reconfigurable and universal SU(2) and U(*N*) operations. We suggested the use of temporal degrees of freedom for reconfigurable universal unitaries in line with recent efforts on the space-time analogy, such as photonic time crystals[20,21] and time disorder[23]. The design of time-coded modulations for high-fidelity QFT and random Haar matrices was demonstrated with superior scalability and the necessary gate number. We can envisage the simultaneous utilization of the space-time degrees of freedom for programmable photonics when reviewing the recent achievements in photonic space-time crystals[42,43].

## Methods

**Low-pass filtering of refractive index modulations.** The time-coded perturbations of the resonant frequencies $\Delta\omega_q(t)$ and the dual-channel gauge fields $\xi^{U,L}_{(p,p+1)}(t)$ are achieved with the refractive index modulations of the parts of the resonators and waveguide loops, respectively. Therefore, although we assume the digital modulation through the linearization of Eq. (2), the obtained modulation cannot be constant as intended due to the response time in the electro-optical[44] or thermal[45] refractive index changes. To reflect the response time, we apply the first-order low-pass filtering to the designed digital signals. For the time-coded modulation signal $f(t)$, the obtained modulation signal $g(t)$ applied to the target optical element is

$$g(t) = \mathcal{F}^{-1}\left[\frac{\omega_c}{\omega_c + i\omega}\mathcal{F}[f(t)]\right], \tag{7}$$

where $\mathcal{F}$ denotes the Fourier transform and $\omega_c$ is the cutoff frequency of the transfer function of the first-order low-pass filter.

**Time-domain analysis of PPTCs.** For the field amplitude $\psi_p$ of the pseudo-spin resonance mode inside the $p$th resonator, we assume the interaction of $\psi_p$ with the incident ($\varphi_p^+$) and scattering waves ($\varphi_p^-$) along the $p$th probe waveguide. Using the Hamiltonian $H(t)$ in Eq. (1), we obtain the following matrix equations:

$$\frac{d}{dt}\Psi(t) = i[H(t) + H_\Delta(t)]\Psi(t) + C_+(t)\Gamma_+(t), \tag{8}$$

$$\Gamma_-(t) = C_0(t)\Gamma_+(t) + C_-(t)\Psi(t), \tag{9}$$

where $\Psi(t) = [\psi_1(t), \psi_2(t), \ldots, \psi_N(t)]^T$ is the time-varying resonator field vector, $\Gamma_\pm(t) = [\varphi_1^\pm(t), \varphi_2^\pm(t), \ldots, \varphi_N^\pm(t)]^T$ are the incident ($\Gamma_+$) and scattering ($\Gamma_-$) field vectors along the probe waveguides, $H_\Delta(t)$ is the perturbing Hamiltonian from the interaction with the probe waveguides, $C_\pm(t)$ are the coupling matrices between the resonators and probe waveguides and $C_0(t)$ is the coupling matrix between the incident and scattering waves along the probe waveguides. We note that $\Gamma_+(t)$ and $H(t)$ are predefined as the input and the result of the decomposition process for the target unitary operation. $C_\pm(t)$ and $C_0(t)$ are also determined by the coupling strength between the resonators and probe waveguides, where $C_0(t)$ is the constant $N \times N$ identity matrix $I_N$ for all cases of our analysis[46]. The goal of the time-domain analysis is to obtain $\Psi(t)$ and $\Gamma_-(t)$ when we know the initial condition $\Psi(0)$.

The $v$-stage Runge-Kutta method[30] derives the numerical iteration for Eq. (8) as follows:

$$\Psi(t+\Delta t) = \Psi(t) + \Delta t \sum_{m=1}^{v} b_m g_m, \qquad (10)$$

where $\Delta t$ is the time-step size, $b_m$ is the Runge-Kutta weights and

$$g_m = i\left[H(t+c_m\Delta t) + H_\Delta(t+c_m\Delta t)\right]\left(\Psi(t) + \Delta t \sum_{n=1}^{m-1} a_{m,n} g_n\right) + C_+(t+c_m\Delta t)\Gamma_+(t+c_m\Delta t), \qquad (11)$$

for the Runge-Kutta nodes $c_m$ and Runge-Kutta matrix elements $a_{m,n}$. The parameters $a_{m,n}$, $b_m$ and $c_m$ are adopted from the seven-stage process for the sixth-order Runge-Kutta method[30].

To evaluate Eqs. (10) and (11), we need to characterize the incidence $\Gamma_+(t)$, the system Hamiltonian $H(t)$ and the interaction parameters with the environment $H_\Delta(t)$ and $C_\pm(t)$, in addition to the initial resonance field $\Psi(0)$. For the predefined incidence $\Gamma_+(t)$, which is discretized with the time step $\Delta t$, we evaluate its value at each stage with the linear interpolation as $\Gamma_+(t + c_m\Delta t) = (1 - c_m) \Gamma_+(t) + c_m\Gamma_+(t + \Delta t)$. On the other hand, the time-varying system Hamiltonian at each stage $H(t + c_m\Delta t)$ is evaluated by using the linearized Eq. (1) with the linearly interpolated physical quantities: $\Delta\omega_p(t + c_m\Delta t) = (1 - c_m)\Delta\omega_p(t) + c_m\Delta\omega_p(t + \Delta t)$ and $\xi_{pq}^{U,L}(t + c_m\Delta t) = (1 - c_m)\xi_{pq}^{U,L}(t) + c_m\xi_{pq}^{U,L}(t + \Delta t)$. In examining the interaction parameters with the environment of PPTCs, we investigate two different cases—isolated and open systems—which support distinct system parameters $H_\Delta(t)$ and $C_\pm(t)$ and initial conditions $\Psi(0)$ and $\Gamma_+(0)$.

In Fig. 3 and Fig. 4a, we explore isolated systems by assuming the absence of the probe waveguides as $H_\Delta(t) = C_\pm(t) = O$, ignoring the initial condition of $\Gamma_+(0)$. Instead, we apply the initial condition of the random resonance field $\psi_p(0) = u[0,A_\psi]\exp(iu[0,2\pi])$, where $u[a,b]$ denotes the uniform random distribution between $a$ and $b$ and $A_\psi$ is set to satisfy the normalization condition of $\Psi(0)$, as $\Sigma_p|\psi_p(0)|^2 = 1$.

In Fig. 4c, we explore open systems interacting with the probe waveguide. Therefore, we assume the initially zero resonator field $\Psi(0) = O$ and the impulse incidence along the probe waveguide $\varphi_p^+(0) = u[0,A_\varphi]\exp(iu[0,2\pi])$ and $\varphi_p^+(t\neq 0) = 0$, where $A_\varphi$ is set to satisfy the normalization condition of $\Gamma_+(0)$. When we assume the weak perturbation of the system in the regime of temporal coupled mode theory[34,47], $C_\pm(t)$ is set to be constant and reciprocal, as $C_\pm(t) = (1/\tau_e)^{1/2}I_N$. For energy conservation, the perturbing Hamiltonian is also constant, as $H_\Delta(t) = iI_N/(2\tau_e)$, reflecting the field decay to the probe waveguides[34].

Using the listed system parameters and $\Psi(0)$ with Eqs. (10) and (11), we calculate the time-varying resonance field $\{\Psi(s\Delta t) \mid s = 1, 2, …, S\}$, where the integer $S$ determines the temporal range of interest for the target unitary operation. The time-varying scattering field $\{\Gamma_-(s\Delta t) \mid s = 1, 2, …, S\}$ to the probe waveguide is directly obtained with Eq. (9).

**Stochastic model for PPTC unitary operations.** To evaluate the fidelity of the effective matrix operation of the PPTC, which is governed by the time-varying, nonlinear and non-Hermitian Hamiltonian, we develop the stochastic model. First, for the PPTC designed to realize the unitary matrix $U \in U(N)$, we prepare $M$ random initial resonance fields $\{\Psi_m(0) \mid m = 1, 2, …, M\}$, which lead to the corresponding outputs $\{\Psi_m(S\Delta t) \mid m = 1, 2, …, M\}$ through the time-domain analysis. Among $M$ pairs of the input and output, we apply the $L$ samplings of the $N$ pairs uniformly at random, without replacement, as $\Psi_n^l(0) \in \{\Psi_m(0)\}$ and $\Psi_n^l(S\Delta t) \in \{\Psi_m(S\Delta t)\}$. At the $l$th sampling ($l = 1, 2, …, L$), we compose the input and output matrices $W_I^l = [\Psi_1^l(0), \Psi_2^l(0), …, \Psi_N^l(0)]$ and $W_O^l = [\Psi_1^l(S\Delta t), \Psi_2^l(S\Delta t), …, \Psi_N^l(S\Delta t)]$. Because the ideal operation is $W_O^l = UW_I^l$, the effective matrix operation $V^l$ for the $l$th sampling is defined as $V^l = W_O^l(W_I^l)^{-1}$. By reflecting the statistical contributions from $V^l$, the fidelity of the PPTC is obtained as $\langle F(U,V^l)\rangle_l$, which leads to $F^{QFT} \equiv \langle F(U^{QFT},V^{QFT,l})\rangle_l$ and $F_k^{Haar} \equiv \langle F(U_k^{Haar},V_k^{Haar,l})\rangle_l$. The conditions of $M \gg N$ and $L \gg N$ guarantee statistical reliability of the stochastic model.

## Data availability

Data used in the current study are available from the corresponding authors upon request.

## Code availability

Codes used in this work will be made available upon request.

(2021).

17. Hennessy, J. L. & Patterson, D. A. *Computer architecture: a quantitative approach* (Elsevier, 2011).

18. Arute, F., Arya, K., Babbush, R., Bacon, D., Bardin, J. C., Barends, R., Biswas, R., Boixo, S., Brandao, F. G. & Buell, D. A. Quantum supremacy using a programmable superconducting processor. *Nature* **574**, 505-510 (2019).

19. Holtmaat, A. & Svoboda, K. Experience-dependent structural synaptic plasticity in the mammalian brain. *Nature Reviews Neuroscience* **10**, 647-658 (2009).

20. Lustig, E., Sharabi, Y. & Segev, M. Topological aspects of photonic time crystals. *Optica* **5**, 1390-1395 (2018).

21. Lyubarov, M., Lumer, Y., Dikopoltsev, A., Lustig, E., Sharabi, Y. & Segev, M. Amplified emission and lasing in photonic time crystals. *Science* **377**, 425-428 (2022).

22. Sharabi, Y., Lustig, E. & Segev, M. Disordered photonic time crystals. *Phys. Rev. Lett.* **126**, 163902 (2021).

23. Kim, J., Lee, D., Yu, S. & Park, N. Unidirectional scattering with spatial homogeneity using correlated photonic time disorder. *Nat. Phys.*, 1-7 (2023).

24. Tirole, R., Vezzoli, S., Galiffi, E., Robertson, I., Maurice, D., Tilmann, B., Maier, S. A., Pendry, J. B. & Sapienza, R. Double-slit time diffraction at optical frequencies. *Nat. Phys.*, doi:10.1038/s41567-023-01993-w (2023).

25. Hafezi, M., Demler, E. A., Lukin, M. D. & Taylor, J. M. Robust optical delay lines with topological protection. *Nat. Phys.* **7**, 907 (2011).

26. Yu, S., Piao, X. & Park, N. Topological Hyperbolic Lattices. *Phys. Rev. Lett.* **125**, 053901 (2020).
24

## Acknowledgements

We acknowledge financial support from the National Research Foundation of Korea (NRF) through the Basic Research Laboratory (No. 2021R1A4A3032027) and Young Researcher Program (No. 2021R1C1C1005031), all funded by the Korean government. This work was also supported by the BK21 FOUR program of the Education and Research Program for Future ICT Pioneers, Seoul National University in 2023. We also acknowledge an administrative support from SOFT foundry institute.

## Author contributions

X.P., S.Y. and N.P. conceived the project idea. X.P. developed the theory, performed numerical analysis and prepared figures. All the authors discussed the results and wrote the manuscript.

## Competing interests

The authors have no conflicts of interest to declare.

## Additional information

**Correspondence and requests for materials** should be addressed to S.Y. or N.P.

**Figure Legends**

**Fig. 1. PPTCs for universal unitaries with $O(N)$ scalability. a,b,** U(4) implementations using the conventional PPC (**a**) and the PPTC (**b**). Black arrows denote the evolution of optical modes along either the spatial ($x$-) or temporal ($t$-) axis. Each coloured box labelled 'SU2' indicates the SU(2) optical gate for two adjacent channels, while the horizontal length of the box represents the spatial or temporal footprint of the gate. **c,d,** The building blocks for the SU(2) operations between the $n$th and $(n+1)$th channels in the PPC (**c**) and the PPTC (**d**). The PPC building block—the SU(2) space gate—consists of two MZIs (grey lines) and two tunable phase shifters (coloured boxes) that employ propagating modes (**c**). The proposed PPTC building block—the SU(2) time gate—is composed of two resonators (grey circles) coupled via two zero-field waveguide loops (grey curved triangles), which support travelling-wave resonance modes (**d**). Tunable phase shifters are applied to the resonators (yellow boxes; **d**) and waveguides or waveguide loops (red and blue boxes; **c** and **d**). Black arrows in **c** and **d** denote the direction of wave propagations. **e,** An example of the nulling process using the Clements design[9] for the diagonalization of $U_4$. The pairs ($p,q$) (black) and $r,s$ (red and blue) denote the matrix element index and the channels connected through the SU(2) gate for nulling ($p,q$), respectively ($p$, $q$, $r$ and $s$ are the integers among 1, 2, 3 and 4). Coloured arrows and boxes in **a**, **b**, and **e** represent two sub-processes of diagonalization, which will be discussed later.

**Fig. 2. SU(2) time gate for PPTCs. a,** Schematic diagram of the SU(2) time gate composed of two resonators (circles) and two waveguide loops (curved triangles). The shaded boxes represent the tunable phase shifters for the resonance perturbations $\Delta\omega_{m,n}(t)$ and the dual-channel gauge fields $\xi_{mn}^{U,L}(t)$. The inset in **a** represents the PPTC composed of the SU(2) time gates. **b,c,** Two orthogonal system states with the even-parity (**b**) and odd-parity (**c**) gauge fields. Black arrows in **a-c** denote the direction of wave circulations. **d,e** The spinor evolutions on the Bloch sphere for the even-parity states (**d**) with $\xi = 2\pi/3$ (red) and $\xi = \pi/6$ (blue), and the odd-parity state (**e**) with $\Delta\omega = 3\times10^{-4}\omega_0$. In **d** and **e**, the plotted temporal range is from 0 to $0.8\pi\tau$ where the black dots denote the initial states.

**Fig. 3. Two-qubit QFT PPTC. a,** The time-coded modulations of the resonance perturbations $\Delta\omega_q$ and the gauge fields in the waveguide loops $\xi^{U,L}_{(p,p+1)}$ for the two-qubit QFT $U^{QFT}$. The signals

undergo low-pass-filtering with the cut-off frequency $\omega_c = \omega_0 / 100$. The duration time of each temporal cell is determined by the elements of $U^{QFT}$ according to Supplementary Note S6. **b,** The corresponding temporal evolutions of the state vectors (modulus and angle of $\psi_q$) at each step of nulling processes. **c,d,** An example of the input (**c**) and output (**d**) of the two-qubit QFT operations. The resonance fields that represent four qubit states $|00\rangle$, $|01\rangle$, $|10\rangle$ and $|11\rangle$ are expressed in the complex plane. In **d**, the solid and dashed lines describe the PPTC output and the solution, respectively. The lifetime of the resonators is set to be $\tau = 500 \times (2\pi / \omega_0)$.

**Fig. 4. Fidelity and measurement of PPTCs. a,** The fidelities of the PPTCs as functions of the cutoff frequency $\omega_c$ in the low-pass-filtering: $F^{QFT}$ (triangles) and $F^{Harr}$ (circles) with $N = 4$ (orange, 2 qubits) and $N = 8$ (blue, 3 qubits). The inset in **a** presents the range of the fidelity larger than 0.9. Circles and error bars show the average and standard deviation of 20 random realizations at each $\omega_c$, respectively. The numbers of the random inputs for estimating $V^{QFT,l}$ or $V_k^{Haar,l}$ and the number of the random samplings for estimating effective matrix operations are set to be $M = 100$ and $L = 100$, respectively. **b,** Schematic diagram of the measurement setup for the PPTCs, employing probe waveguides (blue) to resonators. Red arrows denote the incident ($\varphi_n^+$) and scattering ($\varphi_n^-$) waves. **c,** Scattering power as a function of the lifetime $\tau_e$ to each probe waveguide. All the other parameters are the same as those in Fig. 3 and Methods.

**Supplementary Information for "Programmable photonic time circuits for highly scalable universal unitaries"**

Xianji Piao[1], Sunkyu Yu[2†] and Namkyoo Park[1*]

[1]Photonic Systems Laboratory, Department of Electrical and Computer Engineering, Seoul National University, Seoul 08826, Korea

[2]Intelligent Wave Systems Laboratory, Department of Electrical and Computer Engineering, Seoul National University, Seoul 08826, Korea

E-mail address for correspondence: [†]sunkyu.yu@snu.ac.kr, [*]nkpark@snu.ac.kr

**Note S1. Tight-binding Hamiltonian for coupled resonator lattices**

**Note S2. The spinor Hamiltonian of a SU(2) time gate**

**Note S3. Larmor precession of a spinor**

**Note S4. Universal rotations with SU(2) gates**

**Note S5. Rotation operators of parity states**

**Note S6. Nulling process for universal unitaries**

**Note S7. Fidelity with the probe waveguides**

## Note S1. Tight-binding Hamiltonian for coupled resonator lattices

In this note, we derive the tight-binding Hamiltonian of the coupled resonator lattice, the neighbouring resonators of which are coupled through dual-channel gauge fields (Fig. S1). Consider the coupling of two travelling-wave resonators, each of which supports pseudo-spin modes[1] ($\sigma = \pm 1$). The resonators are coupled indirectly through two waveguide loops that are evanescently coupled to the resonators. The positions of the waveguide loops are set to assign the decay rate of both pseudo-spin modes to be $1/\tau$ for a waveguide loop. For the diatomic unit composed of the $m$th and $n$th resonators, the temporal coupled mode equation for a pseudo-spin mode (e.g., clockwise rotation $\sigma = +1$ in Fig. S1) is[2]

$$\frac{d}{dt}\begin{bmatrix}\psi_m \\ \psi_n\end{bmatrix} = \begin{bmatrix} i[\omega_0 + \Delta\omega_m(t)] - \frac{1}{\tau} & 0 \\ 0 & i[\omega_0 + \Delta\omega_n(t)] - \frac{1}{\tau} \end{bmatrix}\begin{bmatrix}\psi_m \\ \psi_n\end{bmatrix} + \sqrt{\frac{1}{\tau}}\left(\begin{bmatrix}\mu_{mI}^{U} \\ \mu_{nI}^{U}\end{bmatrix} + \begin{bmatrix}\mu_{mI}^{L} \\ \mu_{nI}^{L}\end{bmatrix}\right), \quad (S1)$$

$$\begin{bmatrix}\mu_{mO}^{U} \\ \mu_{nO}^{U}\end{bmatrix} = \begin{bmatrix}\mu_{mI}^{U} \\ \mu_{nI}^{U}\end{bmatrix} - \sqrt{\frac{1}{\tau}}\begin{bmatrix}\psi_m \\ \psi_n\end{bmatrix}, \quad \begin{bmatrix}\mu_{mO}^{L} \\ \mu_{nO}^{L}\end{bmatrix} = \begin{bmatrix}\mu_{mI}^{L} \\ \mu_{nI}^{L}\end{bmatrix} - \sqrt{\frac{1}{\tau}}\begin{bmatrix}\psi_m \\ \psi_n\end{bmatrix}, \quad (S2)$$

$$\begin{bmatrix}\mu_{mI}^{U} \\ \mu_{nI}^{U}\end{bmatrix} = \begin{bmatrix}0 & e^{-i\Phi_{mn}^{U}(t)} \\ e^{-i\Phi_{nm}^{U}(t)} & 0\end{bmatrix}\begin{bmatrix}\mu_{mO}^{U} \\ \mu_{nO}^{U}\end{bmatrix}, \quad \begin{bmatrix}\mu_{mI}^{L} \\ \mu_{nI}^{L}\end{bmatrix} = \begin{bmatrix}0 & e^{-i\Phi_{mn}^{L}(t)} \\ e^{-i\Phi_{nm}^{L}(t)} & 0\end{bmatrix}\begin{bmatrix}\mu_{mO}^{L} \\ \mu_{nO}^{L}\end{bmatrix}, \quad (S3)$$

where $\psi_p$ is the field amplitude of the pseudo-spin resonance mode inside the $p$th resonator, $\mu_{mI}^{U,L}$, $\mu_{mO}^{U,L}$, $\mu_{nI}^{U,L}$, and $\mu_{nO}^{U,L}$ are the field amplitudes at each position of the upper (U) or lower (L) waveguide loops, $\omega_0$ is the reference resonant frequency, $\Delta\omega_p$ is the dynamical perturbation of the resonance frequency of the $p$th resonator, and $\Phi_{pq}^{U,L}(t)$ is the time-varying phase evolution from the $q$th to $p$th resonators along the upper (U) or lower (L) waveguide.

Using Eqs. (S2) and (S3), the incident fields to the resonators $\mu_{mI}^{U,L}$ and $\mu_{nI}^{U,L}$ can be expressed with the resonance fields, as

$$\begin{bmatrix} \mu_{mI}^{U} \\ \mu_{nI}^{U} \end{bmatrix} = \frac{\sqrt{1/\tau}}{1-e^{i\left[\Phi_{mn}^{U}(t)+\Phi_{nm}^{U}(t)\right]}} \begin{bmatrix} 1 & e^{i\Phi_{nm}^{U}(t)} \\ e^{i\Phi_{mn}^{U}(t)} & 1 \end{bmatrix} \begin{bmatrix} \psi_m \\ \psi_n \end{bmatrix},$$

$$\begin{bmatrix} \mu_{mI}^{L} \\ \mu_{nI}^{L} \end{bmatrix} = \frac{\sqrt{1/\tau}}{1-e^{i\left[\Phi_{mn}^{L}(t)+\Phi_{nm}^{L}(t)\right]}} \begin{bmatrix} 1 & e^{i\Phi_{nm}^{L}(t)} \\ e^{i\Phi_{mn}^{L}(t)} & 1 \end{bmatrix} \begin{bmatrix} \psi_m \\ \psi_n \end{bmatrix}.$$

(S4)

By substituting Eq. (S4) into Eq. (S1), we obtain the following equation:

$$\frac{d}{dt}\begin{bmatrix} \psi_m \\ \psi_n \end{bmatrix} = \begin{bmatrix} i[\omega_0 + \Delta\omega_m(t)] - \frac{1}{\tau} & 0 \\ 0 & i[\omega_0 + \Delta\omega_n(t)] - \frac{1}{\tau} \end{bmatrix} \begin{bmatrix} \psi_m \\ \psi_n \end{bmatrix}$$
$$+ \frac{1}{\tau} \frac{1}{1-e^{i\left[\Phi_{mn}^{U}(t)+\Phi_{nm}^{U}(t)\right]}} \begin{bmatrix} 1 & e^{i\Phi_{nm}^{U}(t)} \\ e^{i\Phi_{mn}^{U}(t)} & 1 \end{bmatrix} \begin{bmatrix} \psi_m \\ \psi_n \end{bmatrix}$$
$$+ \frac{1}{\tau} \frac{1}{1-e^{i\left[\Phi_{mn}^{L}(t)+\Phi_{nm}^{L}(t)\right]}} \begin{bmatrix} 1 & e^{i\Phi_{nm}^{L}(t)} \\ e^{i\Phi_{mn}^{L}(t)} & 1 \end{bmatrix} \begin{bmatrix} \psi_m \\ \psi_n \end{bmatrix}.$$

(S5)

To suppress the fields inside the waveguide loops, we preserve the entire phase evolution along each waveguide loop as $\Phi_{mn}^{U,L}(t) + \Phi_{nm}^{U,L}(t) = (2q+1)\pi$ ($q = 0, 1, 2, \ldots$). For the constructive interference between the resonator fields through the evolution along each arm of a waveguide loop, we set each phase evolution to be $\Phi_{mn}^{U,L}(t) = 2q_{mn}^{U,L}\pi + \pi/2 + \xi_{mn}^{U,L}(t)$ and $\Phi_{nm}^{U,L}(t) = 2q_{nm}^{U,L}\pi + \pi/2 - \xi_{mn}^{U,L}(t)$, where $q_{mn}^{U,L}$ and $q_{nm}^{U,L}$ are nonnegative integers. With the assigned $\Phi_{mn}^{U,L}(t)$ and $\Phi_{nm}^{U,L}(t)$, Eq. (S5) becomes:

$$\frac{d}{dt}\begin{bmatrix} \psi_m \\ \psi_n \end{bmatrix} = i\begin{bmatrix} \omega_0 + \Delta\omega_m(t) & \frac{1}{2\tau}\left[e^{-i\xi_{mn}^{U}(t)} + e^{-i\xi_{mn}^{L}(t)}\right] \\ \frac{1}{2\tau}\left[e^{+i\xi_{mn}^{U}(t)} + e^{+i\xi_{mn}^{L}(t)}\right] & \omega_0 + \Delta\omega_n(t) \end{bmatrix} \begin{bmatrix} \psi_m \\ \psi_n \end{bmatrix}.$$

(S6)

Notably, the parameters $q_{mn}^{U,L}$ and $q_{nm}^{U,L}$ can be freely controlled to obtain the practically accessible hardware design of the model of Eq. (S6).

By considering both pseudo-spin modes ($\sigma = \pm 1$) and introducing the creation ($a_{m\sigma}^{\dagger}$) and annihilation ($a_{m\sigma}$) operators for the $\sigma$ pseudo-spin mode of the $m$th resonator, Eq. (S6) derives the dynamical tight-binding Hamiltonian[1] $H = H_{\sigma=+1} + H_{\sigma=-1}$, where

$$H_\sigma = -\omega_0 \left( a_{m\sigma}^\dagger a_{m\sigma} + a_{n\sigma}^\dagger a_{n\sigma} \right)$$
$$- \Delta\omega_m(t) a_{m\sigma}^\dagger a_{m\sigma} - \Delta\omega_n(t) a_{n\sigma}^\dagger a_{n\sigma} \quad (S7)$$
$$- \frac{1}{2\tau}\left[ \left( e^{-i\sigma\xi_{mn}^{U}(t)} + e^{-i\sigma\xi_{mn}^{L}(t)} \right) a_{m\sigma}^\dagger a_{n\sigma} + \text{H.c.} \right],$$

and H.c. denotes the Hermitian conjugate. For the lattice configuration, Eq. (S7) leads to Eq. (1) in the main text.

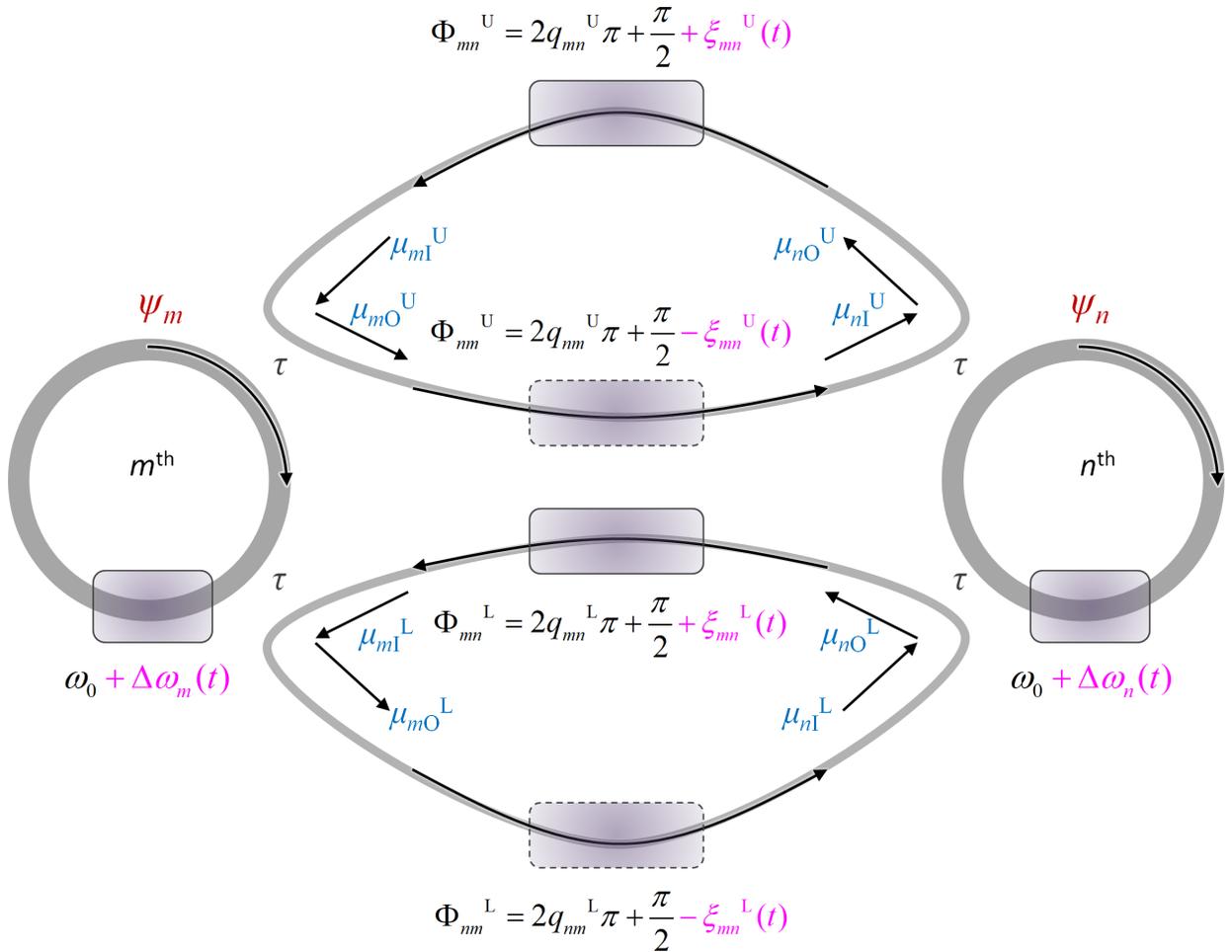

**Fig. S1. Dual-channel coupled resonator unit for SU(2) time gates.** A schematic for the temporal coupled mode theory that describes the indirect couplings between the resonators through dual-channel waveguide loops. The shaded boxes represent the regions of refractive index modulation for time-varying parameters $\Delta\omega_m$, $\Delta\omega_n$, $\pm\xi_{mn}^{U}(t)$, and $\pm\xi_{mn}^{L}(t)$.

**Note S2. The spinor Hamiltonian of a SU(2) time gate**

For the SU(2) gate in Fig. 2a in the main text, we explore the allowed SU(2) operations between the $m$th and $n$th resonators, setting time-varying system parameters as $\xi_{mn}^{U}(t) = \xi^{U}(t)$, $\xi_{mn}^{L}(t) = \xi^{L}(t)$, $\Delta\omega_m(t) = +\Delta\omega(t)$, and $\Delta\omega_n(t) = -\Delta\omega(t)$. When we define the two-level spinor state $\Psi = [\psi_m, \psi_n]^{T}$, Eq. (S6) can be rewritten as:

$$i\frac{d}{dt}\Psi = -\omega_0 I\Psi - \Delta\omega(t)\begin{bmatrix} 1 & 0 \\ 0 & -1 \end{bmatrix}\Psi \\ -\frac{1}{2\tau}\left[\cos\xi^{U}(t) + \cos\xi^{L}(t)\right]\begin{bmatrix} 0 & 1 \\ 1 & 0 \end{bmatrix}\Psi - \frac{1}{2\tau}\left[\sin\xi^{U}(t) + \sin\xi^{L}(t)\right]\begin{bmatrix} 0 & -i \\ i & 0 \end{bmatrix}\Psi. \tag{S8}$$

Using the definition of Pauli matrices, we derive Eq. (2) in the main text.

**Note S3. Larmor precession of a spinor**

We assume the linearized Hamiltonian $H_S$ by assigning the constant values to time-varying system parameters, as $\xi^{U,L}(t) = \xi^{U,L}$ and $\Delta\omega(t) = \Delta\omega$, which leads to the following equation from Eq. (2) in the main text:

$$H_S = -\omega_0 \sigma_0 - \frac{1}{2\tau}\left(\cos\xi^U + \cos\xi^L\right)\sigma_x - \frac{1}{2\tau}\left(\sin\xi^U + \sin\xi^L\right)\sigma_y - \Delta\omega\sigma_z, \tag{S9}$$

where $\sigma_0$ and $\sigma_{x,y,z}$ are the identity matrix and Pauli matrices, respectively. The spinor state is expressed with the Stokes parameters[3] $S_j = \Psi^\dagger \sigma_j \Psi$ ($j = 0, x, y,$ and $z$). By employing the linear equation $id\Psi/dt = H_S\Psi$, we can achieve the geometrical description of the spinor evolution in the form of the equation of motion for the spinor state[4-6]. First, the evolution of $S_j$ is

$$\frac{dS_j}{dt} = \frac{d}{dt}\left(\Psi^\dagger \sigma_j \Psi\right) = \frac{d\Psi^\dagger}{dt}\sigma_j \Psi + \Psi^\dagger \sigma_j \frac{d\Psi}{dt}. \tag{S10}$$

We apply $d\Psi/dt = -iH_S\Psi$ and $d\Psi^\dagger/dt = i\Psi^\dagger H_S$ due to the Hermiticity of $H_S$, which gives

$$\frac{dS_j}{dt} = i\Psi^\dagger\left(H_S\sigma_j - \sigma_j H_S\right)\Psi = i\Psi^\dagger\left[H_S, \sigma_j\right]\Psi, \tag{S11}$$

where $[A,B]$ is the commutator of the operators $A$ and $B$. Because $[H_S,\sigma_0] = O$, we get

$$\frac{dS_0}{dt} = 0, \tag{S12}$$

which guarantees the conservation of the Bloch sphere radius for the Hermitian Hamiltonian $H_S$. With the Pauli matrix algebra and Eq. (S9), we achieve the following relations for $S_{x,y,z}$:

$$\begin{aligned}\frac{dS_x}{dt} &= 2\Delta\omega S_y - \frac{1}{\tau}\left(\sin\xi^U + \sin\xi^L\right)S_z, \\ \frac{dS_y}{dt} &= \frac{1}{\tau}\left(\cos\xi^U + \cos\xi^L\right)S_z - 2\Delta\omega S_x, \\ \frac{dS_z}{dt} &= \frac{1}{\tau}\left(\sin\xi^U + \sin\xi^L\right)S_x - \frac{1}{\tau}\left(\cos\xi^U + \cos\xi^L\right)S_y.\end{aligned} \tag{S13}$$

Because the radius of the Bloch sphere is conserved from Eq. (S12), the spinor state can be expressed as the Stokes vector $\mathbf{S} = [S_x, S_y, S_z]^\mathrm{T}$. By defining the pseudo-magnetic field $\mathbf{B}$ shown in Eq. (3) in the main text, the geometrical evolution of $\mathbf{S}$ on the Bloch sphere is described by the Larmor precession equation $d\mathbf{S}/dt = \mathbf{S} \times \mathbf{B}$.

**Note S4. Universal rotations with SU(2) gates**

For the concise form of Eq. (2) in the main text, $H_S = -\omega_0\sigma_0 - \mathbf{B}\cdot\boldsymbol{\sigma}/2$, the corresponding rotation operator $R_\mathbf{B}$ during the time $t$ is achieved with

$$R_\mathbf{B} = \exp(-iH_S t) = \exp(i\omega_0 t)\exp\left(i\frac{\mathbf{B}\cdot\boldsymbol{\sigma}}{2}t\right), \tag{S14}$$

because $\sigma_0$ commutes with the other Pauli matrices $\sigma_{1\text{-}3}$. We employ the property of the matrix exponential[7]:

$$\exp(iAx) = \cos(x)\sigma_0 + i\sin(x)A, \tag{S15}$$

for a real number $x$ and a matrix $A$ such that $A^2 = \sigma_0$. For the pseudo-magnetic field $\mathbf{B} = \mathbf{e}_\mathbf{B} B_0$, the unit vector $\mathbf{e}_\mathbf{B}$ satisfies $(\mathbf{e}_\mathbf{B}\cdot\boldsymbol{\sigma})^2 = \sigma_0$ for $\boldsymbol{\sigma} = \mathbf{e}_x\sigma_x + \mathbf{e}_y\sigma_y + \mathbf{e}_z\sigma_z$. Therefore, we achieve

$$\exp\left(i\frac{\mathbf{B}\cdot\boldsymbol{\sigma}}{2}t\right) = \exp\left(i\mathbf{e}_\mathbf{B}\cdot\boldsymbol{\sigma}\frac{B_0 t}{2}\right) = \cos\left(\frac{B_0 t}{2}\right)\sigma_0 + i\sin\left(\frac{B_0 t}{2}\right)\mathbf{e}_\mathbf{B}\cdot\boldsymbol{\sigma}, \tag{S16}$$

which results in Eq. (4) in the main text.

**Note S5. Rotation operators of parity states**

For the even-parity state from $\xi^U = \xi^L = \xi$ and $\Delta\omega = 0$ (Fig. 2b in the main text), the rotation operator $U_2(t)$ from Eq. (4) in the main text is

$$U_2(t) = e^{i\omega_0 t} \begin{bmatrix} \cos\dfrac{t}{\tau} & ie^{-i\xi}\sin\dfrac{t}{\tau} \\ ie^{+i\xi}\sin\dfrac{t}{\tau} & \cos\dfrac{t}{\tau} \end{bmatrix}, \quad (S17)$$

where $t$ determines the amount of the rotation of a spinor state about the pseudo-magnetic field on the $xy$ plane. The direction of the pseudo-magnetic field is determined by the angle $\xi$, as shown in Eq. (3) in the main text.

For the odd-parity state with quarter-wave gauge fields from $\xi^U = -\xi^L = \pi/2$, while allowing for $\Delta\omega \neq 0$, the rotation operator $U_2(t)$ from Eq. (4) in the main text is

$$U_2(t) = e^{i\omega_0 t} \begin{bmatrix} e^{+i\Delta\omega t} & 0 \\ 0 & e^{-i\Delta\omega t} \end{bmatrix}, \quad (S18)$$

where $t$ determines the amount of the rotation of a spinor state about the pseudo-magnetic field along the $z$-axis. When $\Delta\omega = 0$, the spinor state is preserved except for the global phase evolution.

**Note S6. Nulling process for universal unitaries**

To decompose an *N*-dimensional unitary matrix $U_N$ for its realization with a PPTC, we extend the Clements design[8] to the coupled resonator lattices examined in Supplementary Notes S1-S5. The Clements design utilizes the sequential nulling of the pairs of super- and sub-diagonals of $U_N$. A unit nulling process is achieved by multiplying the *N*-dimensional unitary matrix $(T_m^l)^\dagger$ to *U*, where $T_m^l$ leads to the SU(2) operation between the *m*th and (*m*+1)th channels. The entire nulling process is composed of a series of two alternating processes: the (*l*, *m*) element nulling with $U(T_m^l)^\dagger$ and the (*m*+1, *l*) element nulling with $(T_m^l)^\dagger U$.

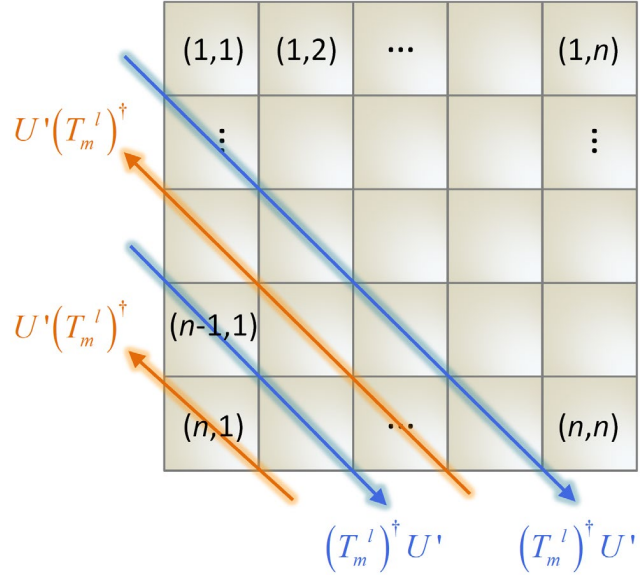

**Fig. S2. The sequence of nulling processes for PPTCs.** The nulling process applied to *U'* aims to set the target element of *U'* zero. The process is composed of a series of two sub-processes: the process $U'(T_m^l)^\dagger$ for the nulling of the (*l*, *m*) element (orange arrows) and the process $(T_m^l)^\dagger U'$ for the nulling of the (*m*+1, *l*) element (blue arrows).

For the SU(2) operation between the *m*th and (*m*+1)th channels, the $N \times N$ matrix $T_m^l$ has the form of

$$T_m^l = \begin{bmatrix} 1 & 0 & & \cdots & & 0 & 0 \\ 0 & \ddots & & \cdots & & & 0 \\ & & t_{m,m}^l & t_{m,m+1}^l & & & \\ \vdots & \vdots & t_{m+1,m}^l & t_{m+1,m+1}^l & & \vdots & \\ 0 & & & & \ddots & & 0 \\ 0 & 0 & & \cdots & & 0 & 1 \end{bmatrix}, \quad (S19)$$

where $T_m^l$ follows the $N \times N$ identity matrix except for the block matrix between the $m$th and $(m+1)$th channels, and the matrix elements $t_{p,q}^l$ ($p$, $q = m$ or $m+1$) lead to the SU(2) operation defined by the rotation operator $R_\mathbf{B}(t)$ in Eq. (4) in the main text. The goal of the nulling process is to calculate the necessary parameters of $\xi^{U,L}$, $\Delta\omega$, and the duration time $t$ of the SU(2) time gate for each $T_m^l$ that results in the nulling of the target element in $U_N$.

When we define the transitional unitary matrix as $U'$ during the nulling process of $U_N$, the $U'(T_m^l)^\dagger$ nulling process leads to:

$$\begin{aligned} U'\left(T_m^l\right)^\dagger & \\ &= \begin{bmatrix} U'_{(1,1)} & U'_{(1,2)} & \cdots & U'_{(1,n)} \\ U'_{(2,1)} & U'_{(2,2)} & \cdots & U'_{(2,n)} \\ \vdots & \vdots & & \vdots \\ U'_{(n,1)} & U'_{(n,2)} & \cdots & U'_{(n,n)} \end{bmatrix} \begin{bmatrix} 1 & 0 & & \cdots & & 0 & 0 \\ 0 & \ddots & & \cdots & & & 0 \\ & & \left(t_{m,m}^l\right)^* & \left(t_{m+1,m}^l\right)^* & & & \\ \vdots & \vdots & \left(t_{m,m+1}^l\right)^* & \left(t_{m+1,m+1}^l\right)^* & & \vdots & \\ 0 & & & & \ddots & & 0 \\ 0 & 0 & & \cdots & & 0 & 1 \end{bmatrix} \\ &= \begin{bmatrix} U'_{(p,1)} & \cdots & U'_{(p,m-1)} & V_{(p,m)} & V_{(p,m+1)} & U'_{(p,m+2)} & \cdots & U'_{(p,n)} \end{bmatrix} \quad (1 \leq p \leq n), \end{aligned} \quad (S20)$$

where $U'_{(p,k)}$ ($1 \leq p \leq n$) denotes the $k$th column of the $U'$ matrix, and $V_{(p,m)}$ and $V_{(p,m+1)}$ represent the SU(2)-transformed states, as

$$V_{(p,m)} = U'_{(p,m)}\left(t_{m,m}^{\ l}\right)^* + U'_{(p,m+1)}\left(t_{m,m+1}^{\ l}\right)^*,$$
$$V_{(p,m+1)} = U'_{(p,m)}\left(t_{m+1,m}^{\ l}\right)^* + U'_{(p,m+1)}\left(t_{m+1,m+1}^{\ l}\right)^*. \quad \text{(S21)}$$

The nulling of the (*l*, *m*) element with $U'(T_m^l)^\dagger$ requires the following condition:

$$U'_{(l,m)}\left(t_{m,m}^{\ l}\right)^* + U'_{(l,m+1)}\left(t_{m,m+1}^{\ l}\right)^* = 0. \quad \text{(S22)}$$

Similarly, the $(T_m^l)^\dagger U'$ nulling process leads to:

$$(T_m^l)^\dagger U' = \begin{bmatrix} 1 & 0 & & \cdots & & 0 & 0 \\ 0 & \ddots & & \cdots & & & 0 \\ & & \left(t_{m,m}^{\ l}\right)^* & \left(t_{m+1,m}^{\ l}\right)^* & & & \\ \vdots & \vdots & \left(t_{m,m+1}^{\ l}\right)^* & \left(t_{m+1,m+1}^{\ l}\right)^* & & \vdots & \\ 0 & & & & \ddots & & 0 \\ 0 & 0 & & \cdots & & 0 & 1 \end{bmatrix} \begin{bmatrix} U'_{(1,1)} & U'_{(1,2)} & \cdots & U'_{(1,n)} \\ U'_{(2,1)} & U'_{(2,2)} & \cdots & U'_{(2,n)} \\ \vdots & \vdots & & \vdots \\ U'_{(n,1)} & U'_{(n,2)} & \cdots & U'_{(n,n)} \end{bmatrix}$$

$$= \begin{bmatrix} U'_{(1,p)} \\ \vdots \\ U'_{(m-1,p)} \\ V_{(m,p)} \\ V_{(m+1,p)} \\ U'_{(m+2,p)} \\ \vdots \\ U'_{(n,p)} \end{bmatrix} \quad (1 \leq p \leq n), \quad \text{(S23)}$$

where $U'_{(k,p)}$ (1 ≤ *p* ≤ *n*) denotes the *k*th row of the $U'$ matrix, and $V_{(m,p)}$ and $V_{(m+1,p)}$ represent the SU(2)-transformed states, as

$$V_{(m,p)} = U'_{(m,p)}\left(t_{m,m}^{\ l}\right)^* + U'_{(m+1,p)}\left(t_{m+1,m}^{\ l}\right)^*,$$
$$V_{(m+1,p)} = U'_{(m,p)}\left(t_{m,m+1}^{\ l}\right)^* + U'_{(m+1,p)}\left(t_{m+1,m+1}^{\ l}\right)^*. \quad \text{(S24)}$$

The nulling of the (*m*+1, *l*) element with $(T_m^l)^\dagger U'$ requires the following condition:

$$U'_{(m,l)}\left(t_{m,m+1}^{l}\right)^* + U'_{(m+1,l)}\left(t_{m+1,m+1}^{l}\right)^* = 0. \tag{S25}$$

As shown in Eqs. (S22) and (S25), the unitary matrix $T_m^l$ should reflect the complex-valued ratio between the *m*th and (*m*+1)th channels while imposing the identity matrix operation on the other channels according to Eq. (S19). We thus employ the even-parity phase between the *m*th and (*m*+1)th channels, which leads to the spinor rotation in Eq. (S17), while applying the odd-parity phase with $\Delta\omega = 0$ to the other pairs of the neighbouring channels using Eq. (S18).

For the SU(2) transformation $U'(T_m^l)^\dagger$, the gauge field $\xi$ and the duration time $t$ for the even-parity phase of the *m*th-(*m*+1)th coupled resonators is achieved with Eq. (S22), as follows:

$$\begin{bmatrix} \text{I. } \dfrac{t}{\tau} = \dfrac{\pi}{2}, \text{ arbitrary } \xi(=0) \quad \left(\text{when } U'_{(l,m)} \neq 0, \ U'_{(l,m+1)} = 0\right), \\ \text{II. } \dfrac{t}{\tau} = 0, \text{ arbitrary } \xi(=0) \quad \left(\text{when } U'_{(l,m)} = 0, \ U'_{(l,m+1)} \neq 0\right), \\ \text{III. arbitrary } \dfrac{t}{\tau}(=0), \text{ arbitrary } \xi(=0) \quad \left(\text{when } U'_{(l,m)} = 0, \ U'_{(l,m+1)} = 0\right), \\ \text{IV. } \dfrac{t}{\tau} = \arctan\left|\dfrac{U'_{(l,m)}}{U'_{(l,m+1)}}\right|, \ \xi = \angle\dfrac{U'_{(l,m)}}{U'_{(l,m+1)}} - \dfrac{\pi}{2} \quad \left(\text{when } U'_{(l,m)} \neq 0, \ U'_{(l,m+1)} \neq 0\right). \end{bmatrix} \tag{S26}$$

For the SU(2) transformation $(T_m^l)^\dagger U'$, $\xi$ and $t$ for the even-parity phase of the *m*th-(*m*+1)th coupled resonators is achieved with Eq. (S25), as follows:

$$\begin{bmatrix}
\text{I. } \dfrac{t}{\tau} = \dfrac{\pi}{2}, \text{ arbitrary } \xi(=0) \quad \left(\text{when } U'_{(m+1,l)} \neq 0,\ U'_{(m,l)} = 0\right), \\
\text{II. } \dfrac{t}{\tau} = 0, \text{ arbitrary } \xi(=0) \quad \left(\text{when } U'_{(m+1,l)} = 0,\ U'_{(m,l)} \neq 0\right), \\
\text{III. arbitrary } \dfrac{t}{\tau}(=0), \text{ arbitrary } \xi(=0) \quad \left(\text{when } U'_{(m+1,l)} = 0,\ U'_{(m,l)} = 0\right), \\
\text{IV. } \dfrac{t}{\tau} = \arctan\left|\dfrac{U'_{(m+1,l)}}{U'_{(m,l)}}\right|,\ \xi = \angle\dfrac{U'_{(m+1,l)}}{U'_{(m,l)}} - \dfrac{\pi}{2} \quad \left(\text{when } U'_{(m+1,l)} \neq 0,\ U'_{(m,l)} \neq 0\right).
\end{bmatrix} \quad (S27)$$

The identity matrix operation with the decoupling state between the other channels except for the pair of the $m$th and $(m+1)$th channels is achieved by the odd-parity phase without resonance perturbations, as $\xi = \pi/2$ and $\Delta\omega = 0$. The duration time for the decoupling state is set to be identical to the obtained $t$ in Eqs. (S26) and (S27).

After the entire nulling processes of the Clements design, we obtain the diagonal matrix $D$, as follows:

$$D = \left[\prod_{(l_F, m_F)} \left(T_{m_F}^{l_F}\right)^\dagger\right] U_N \left[\prod_{(l_B, m_B)} \left(T_{m_B}^{l_B}\right)^\dagger\right], \quad (S28)$$

where $l_F$ and $m_F$ (or $l_B$ and $m_B$) denote the indices of $T_m^l$ for the $(T_m^l)^\dagger U'$ (or $U'(T_m^l)^\dagger$) nulling operation. The sequences of the pairs $(l_F, m_F)$ and $(l_B, m_B)$ are determined by the nulling process described in Fig. S2. The result of the $U_N$ decomposition becomes Eq. (5) in the main text with the reversed sequences of the pairs $(l_F, m_F)'$ and $(l_B, m_B)'$ of the original ones $(l_F, m_F)$ and $(l_B, m_B)$.

The diagonal matrix $D$ can be realized with the odd-parity phase with resonance perturbations, as $\xi = \pi/2$ and $\Delta\omega_m = \angle d_m$, where $d_m$ is the $m$th diagonal element of $D$ and $\Delta\omega_m$ is the following resonance perturbation of the $m$th resonator.

In the Clements design using the SU(2) gates composed of Mach-Zehnder interferometers and phase shifters, the transformation of $TD = D'T'^\dagger$ is employed[8]. However, the proposed SU(2) time gates do not allow such a transformation. Furthermore, due to different values of $t$ for each $T_m^l$ as shown in Eqs. (S26) and (S27), the simultaneous operation of different $T_m^l$ matrices is not straightforward in terms of the synchronization of each resonator element for $U(N)$ operations. Therefore, the circuit implementation of $U_N$ is achieved directly through the cascaded temporal dynamics of unit unitary and diagonal operations based on Eq. (5) in the main text.

**Note S7. Fidelity with the probe waveguides**

When we assume temporally bound incident pulses from $t = 0$ to $t = t_{src}$, as $\Gamma_+(t > t_{src}) = O$, Eqs. (7) and (8) in the main text at $t > t_{src}$ become

$$\frac{d}{dt}\Psi(t) = i[H(t) + H_\Delta(t)]\Psi(t), \tag{S29}$$

$$\Gamma_-(t) = C_-(t)\Psi(t). \tag{S30}$$

We note that $C_-(t) = (1/\tau_e)^{1/2} I_N$ and $H_\Delta(t) = iI_N/(2\tau_e)$. Therefore, in estimating $\Psi(t)$, Eq. (S29) is equivalent to that of the isolated system in Figs. 3 and 4a in the main text, except for the decay $\exp(-1/(2\tau_e))$, which contributes equally to all of the resonators. With the normalization of the output power and sufficiently small $t_{src}$, the probe waveguides that are coupled equally to the resonators do not affect the fidelity of the PPTC. Furthermore, from Eq. (S30), the scattering power changes with $\exp(-1/\tau_e)$, which is solely determined by $\tau_e$.